\documentclass[aps,showpacs,superscriptaddress,12pt]{revtex4-2}

\usepackage[english]{babel}

\usepackage[letterpaper,top=2cm,bottom=2cm,left=3cm,right=3cm,marginparwidth=1.75cm]{geometry}

\usepackage{amsmath}
\usepackage{mathrsfs}
\usepackage{graphicx}
\usepackage{subfigure}
\usepackage{multirow}
\usepackage[colorlinks=true, allcolors=blue]{hyperref}
\usepackage{float}
\usepackage{fancyhdr}

\newcommand{\jpsi}{J/\psi}

\newcommand{\GeVcc}{~\text{GeV}/c^2}
\newcommand{\MeV}{~\text{MeV}}
\newcommand{\MeVcc}{~\text{MeV}/c^2}

\fancypagestyle{firstpage}{
  \fancyhf{}
  \rhead{\footnotesize Submitted to \textit{Chinese Physics C}}
  
}

\begin{document}
\title{\boldmath
Enhanced evidence of $X(7200)$ and improved measurements of $X(6900)$ parameters from a combined LHCb-ATLAS-CMS analysis }

\author{Yuan Wang}
\affiliation{Department of Physics, Hebei University, Baoding 071002, China}
\affiliation{Hebei Key Laboratory of High-precision Computation and Application of Quantum Field Theory, Baoding 071002, China}
\affiliation{Hebei Research Center of the Basic Discipline for Computational Physics, Baoding 071002, China}
\author{Ran Li}
\affiliation{Department of Physics, Hebei University, Baoding 071002, China}
\affiliation{Hebei Key Laboratory of High-precision Computation and Application of Quantum Field Theory, Baoding 071002, China}
\affiliation{Hebei Research Center of the Basic Discipline for Computational Physics, Baoding 071002, China}
\author{Bin Zhong}
\email{zhongb@njnu.edu.cn (Corresponding Author)}
\affiliation{Department of Physics and Institute of Theoretical Physics, Nanjing Normal University, Nanjing 210023, China}
\affiliation{Nanjing Key Laboratory of Particle Physics and Astrophysics, Nanjing 210023, China}

\author{Ya-Qian Wang}
\email{whyaqm@hbu.edu.cn (Corresponding Author)}
\affiliation{Department of Physics, Hebei University, Baoding 071002, China}
\affiliation{Hebei Key Laboratory of High-precision Computation and Application of Quantum Field Theory, Baoding 071002, China}
\affiliation{Hebei Research Center of the Basic Discipline for Computational Physics, Baoding 071002, China}

\date{\today}

\begin{abstract}
We report stronger evidence for the $X(7200)$ state and markedly improved measurements of the $X(6900)$ resonance parameters based on a combined analysis of the di-$J/\psi$ mass spectrum using published data from LHCb, ATLAS, and CMS. Through simultaneous fits to the datasets from all three experiments, we observe the $X(6900)$ with overwhelming significance ($>12\sigma$) and determine its mass and width with improved precision. For the $X(7200)$, we find consistent signals across multiple interference models, with significances ranging from $3.7\sigma$ to $6.6\sigma$; in the best-fit model (the CMS three-resonance scheme), the significance reaches $6.6\sigma$, providing substantially stronger evidence for this state. Our results underscore the essential role of interference effects in fully charmed tetraquark spectroscopy and offer new constraints on their production mechanisms at the LHC.
\end{abstract}


\maketitle

\thispagestyle{firstpage}

\section{Introduction}

The constituent quark model~\cite{Gell-Mann:1964ewy,Zweig:1964jf}, since its proposal by Gell-Mann and Zweig, has been remarkably successful in classifying hadrons and describing their spectroscopic properties. In the conventional picture, hadrons are categorized as mesons ($q\bar{q}$) and baryons ($qqq$ or $\bar{q}\bar{q}\bar{q}$). However, Quantum Chromodynamics (QCD), the fundamental theory of strong interactions, does not forbid the existence of more complex color-singlet structures. These states—collectively termed exotic hadrons—include glueballs, hybrids, and multiquark configurations (tetraquarks and pentaquarks) and have been the subject of intense theoretical and experimental interest for decades.
The renaissance of hadron spectroscopy in the 21st century, driven by B-factories and the Large Hadron Collider (LHC), has provided compelling evidence for the existence of exotic hadrons. The observation of the $X(3872)$ by the Belle Collaboration~\cite{Belle:2003nnu} marked the beginning of the $XYZ$ era. Since then, numerous charmonium-like and bottomonium-like states, as well as multiquark states, have been reported~\cite{LHCb:2015yax,LHCb:2019kea,LHCb:2020bwg}. Among these candidates, fully heavy tetraquark states, composed of four heavy quarks ($Q\bar{Q}Q\bar{Q}$, where $Q=c, b$), are of particular importance. Unlike $XYZ$ states that contain light quarks, the fully heavy system is free from light-meson exchange and chiral dynamics, offering a pristine environment to probe heavy-quark interactions in the nonperturbative regime of QCD.

The search for fully charmed tetraquarks ($cc\bar{c}\bar{c}$) has primarily focused on the di-$J/\psi$ invariant mass spectrum. In 2020, the LHCb Collaboration reported a breakthrough observation with 9 fb$^{-1}$ of proton-proton collision data collected at center-of-mass energies $\sqrt{s}=$7, 8, and 13 TeV~\cite{LHCb:2020bwg}. A distinct, narrow structure, named $X(6900)$, was observed around $6.9\GeVcc$ with a significance greater than $5\sigma$. Following this discovery, the ATLAS and CMS collaborations confirmed the existence of $X(6900)$ with data taken at $\sqrt{s}=$13 TeV, and also reported a hint of, or evidence for, another structure near 7.2 $\GeVcc$~\cite{ATLAS:2023bft,CMS:2023owd}. The ATLAS paper~\cite{ATLAS:2023bft} reports that the $X(6400)$ (denoted as $X_1$ in our model) and the $X(6600)$ (denoted as $X_2$) appear as broad enhancements near the di-$J/\psi$ threshold, where nonresonant dynamics—such as coupled-channel effects~\cite{Dong:2020nwy}, triangle singularities~\cite{Guo:2019twa}, or Pomeron exchange~\cite{Gong:2020bmg}—may contribute significantly. Consequently, these structures are not conclusively established as genuine resonances, and their interpretation as compact tetraquark states is less favored than that of the narrower $X(6900)$.

While the existence of resonant structures in the di-$J/\psi$ mass spectrum is now well established, the interpretation of these states remains controversial~\cite{Wang:2020wrp,Chen:2020xwe,Bedolla:2019zwg,Jin:2020jfc,liu:2020eha,Liu:2021rtn,Zhu:2020xni,Giron:2020wpx}. The measured masses and widths of $X(6900)$ vary among experiments, owing to differences in background modeling and possible interference effects~\cite{LHCb:2020bwg,ATLAS:2023bft,CMS:2023owd}. Furthermore, the nature of the structures in the higher-mass region, specifically around $7.2\GeVcc$ (hereafter referred to as $X(7200)$), remains less well understood. Considerations of such states date back to 1976~\cite{Iwasaki:1976cn,Chao:1980dv} and have recently attracted renewed interest~\cite{Berezhnoy:2011xn,Wu:2016vtq,Bai:2016int,Wang:2017jtz,Anwar:2017toa,Richard:2017vry,Esposito:2018cwh,Karliner:2016zzc,Liu:2019zuc,Chen:2016jxd,Wang:2019rdo,Wang:2025hex}. However, non-tetraquark interpretations of this structure have also been proposed~\cite{Gong:2020bmg,Wan:2020fsk,Wang:2020tpt}. A recent CMS angular analysis~\cite{CMS:2025fpt} favors $J^{PC}=2^{++}$, challenging molecular models (which typically expect $0^{++}$) while supporting compact tetraquark scenarios.

\section{Data sets}
The LHCb Collaboration analyzed the di-$J/\psi$ invariant-mass spectrum by describing the $X(6900)$ with an independent Breit-Wigner (BW) resonance and modeling the background using two distinct approaches~\cite{LHCb:2020bwg}:
\begin{itemize}
  \item A \textit{coherent fit} modeling an interfering BW resonance with the non-resonant continuum yields $M=6886 \pm 11\pm11$ $\MeVcc$ and $\Gamma = 168 \pm 33\pm69$ $\MeV$.
    \item An \textit{incoherent fit}, treating wide peaks as isolated BW resonances superimposed on the background, yields $M = 6905 \pm 11 \pm 7$ $\MeVcc$ and $\Gamma = 80 \pm 19 \pm 33$ $\MeV$.
\end{itemize}
This dual strategy enabled systematic studies of the resonance parameters and of their model dependencies.

The ATLAS Collaboration analyzed $140~\mathrm{fb}^{-1}$ of $13~\mathrm{TeV}$ $pp$ collision data to study di-$J/\psi$ production using two models~\cite{ATLAS:2023bft}:
\begin{itemize}
    \item \textit{Signal Interference}: The coherent sum of the BW amplitudes for $X(6400)$, $X(6600)$, and $X(6900)$:
    $M=6910\pm10\pm10\MeVcc$, $\Gamma=150\pm30\pm10\MeV$.
    \item \textit{Background Interference}: The single parton scattering (SPS) background interferes with lower-mass resonances, whereas $X(6900)$ remains isolated: $M=6860\pm30^{+10}_{-20}\MeVcc$, $\Gamma=110\pm50^{+20}_{-10}\MeV$.
    \end{itemize}
The observed $\sim 50~\MeVcc$ mass shift and $\sim 40\MeV$ width difference underscore the critical impact of interference modeling on resonance-parameter extraction.

The CMS Collaboration analyzed $135~\mathrm{fb}^{-1}$ of $pp$ collision data and studied the di-$J/\psi$ spectrum using two distinct approaches~\cite{CMS:2023owd}:

\begin{itemize}
    \item \textit{Noninterfering model}: Three isolated BW resonances ($X(6600)$, $X(6900)$, and $X(7200)$) plus a background term $B(m)$
    \item \textit{Interference Model}: A coherent sum of the three resonances with relative phases.
\end{itemize}

The key findings demonstrate strong interference effects:
\begin{itemize}
    \item The $X(6900)$ mass shifts from $6927 \pm 9\pm 4~\MeVcc$ (no interference) to $6847^{+44 \, +48}_{-28 \, -20}$ $\MeVcc$ (with interference).
    \item The width increases from $122^{+24}_{-21}\pm 18$ $\MeV$ to $191^{+66 \, +25}_{-49 \, -17}$ $\MeV$.
    \item The $X(7200)$ parameters exhibit a similar model dependence.
\end{itemize}
The $\sim80~\MeVcc$ mass shift and the factor-of-1.6 change in width highlight the necessity of interference modeling in all-charm spectroscopy.

\section{Simultaneous fits}
We perform a simultaneous fit to the di-$\jpsi$ mass spectra from LHCb, ATLAS, and CMS. To probe the nature of the observed resonances---particularly the $X(6900)$ and $X(7200)$ states---we implement four distinct fit hypotheses. These models incorporate different treatments of the coherent sum of BW amplitudes, reflecting different assumptions about inter-resonance interference. 
 The statistical significance of each state ($X(6900)$ or $X(7200)$) is determined by computing $\Delta\chi^2$ and $\Delta {\rm NDF}$, where $\Delta\chi^2$ is the difference in $\chi^2$ and $\Delta {\rm NDF}$ is the change in degrees of freedom when including or excluding the resonance component in the fit. 
The fit function for each experiment combines signal components—modeled by $S$-wave relativistic BW amplitudes—with a nonresonant background.  
The background comprises two contributions: single-parton scattering (SPS) and double-parton scattering (DPS).  
The SPS component is modeled by an exponential function, denoted $f_{\rm SPS}$, while the DPS component, $f_{\rm DPS}$, is taken directly from the published results~\cite{LHCb:2020bwg,ATLAS:2023bft,CMS:2023owd}. Consequently, $f_{\rm DPS}$ already includes the phase-space factor and is fixed in our fit.

\begin{figure}[!h]
    \centering
    \subfigure[ATLAS.]{
    \includegraphics[width=0.5\textwidth]{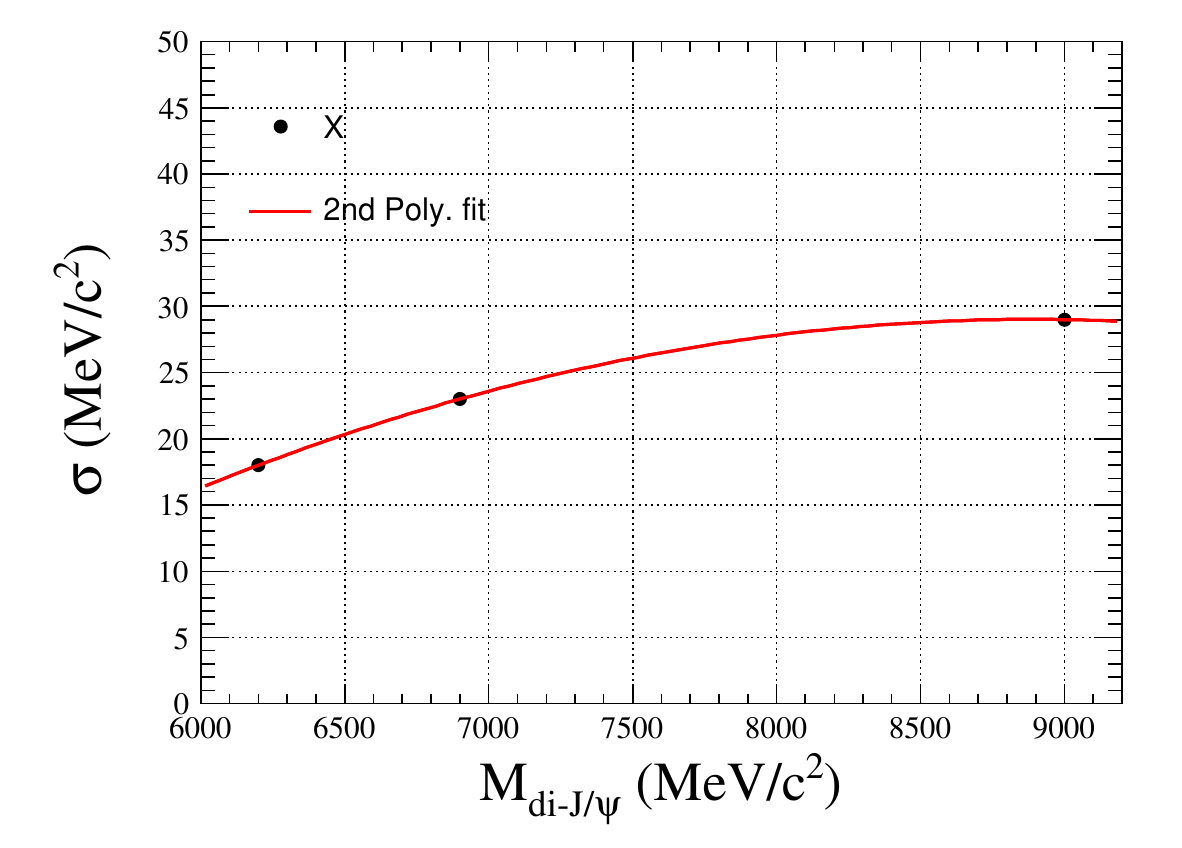}
    }%
    \subfigure[CMS.]{
    \includegraphics[width=0.5\textwidth]{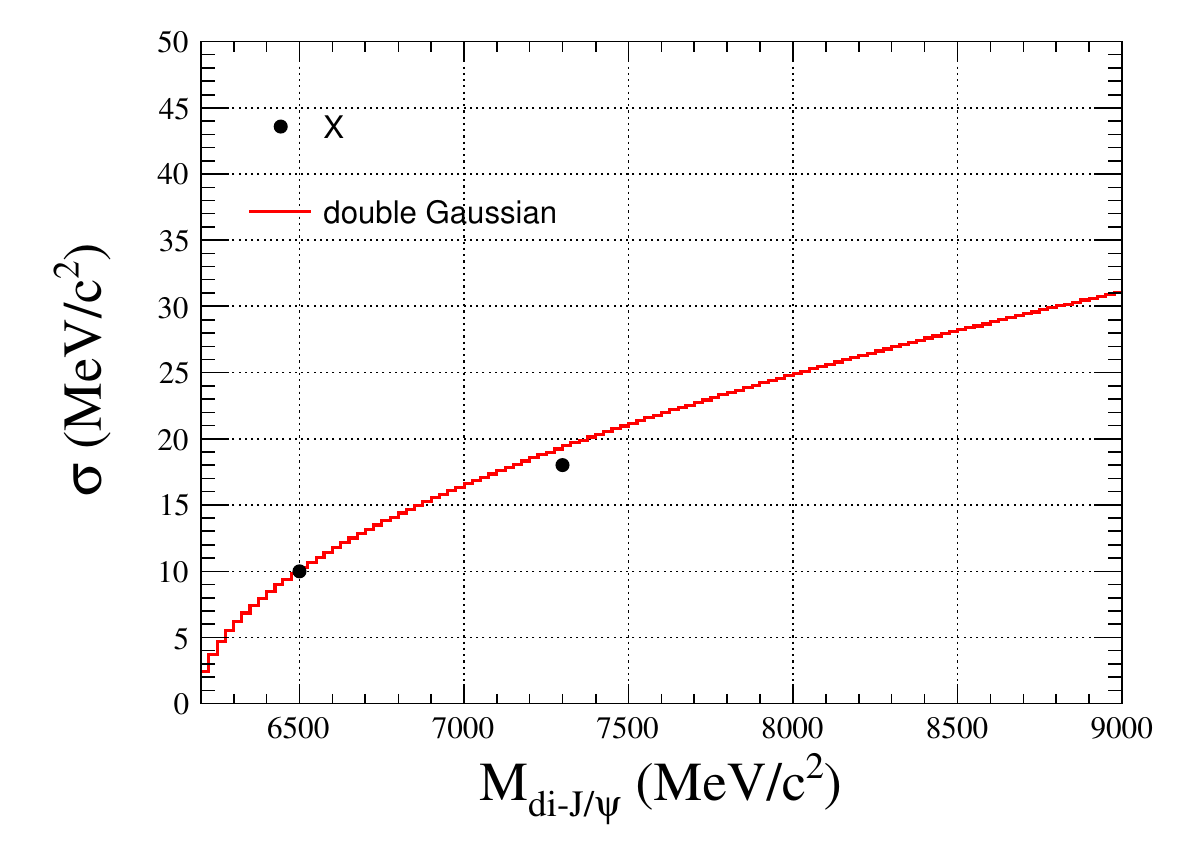}}%
    \caption{Mass resolutions for ATLAS and CMS.}
    \label{f_res}
\end{figure}

The mass resolution is below 5~$\MeVcc$ across the full mass range for LHCb~\cite{LHCb:2020bwg}, while it ranges from 18~$\MeVcc$ to 29~$\MeVcc$ (with an average of 23~$\MeVcc$) for ATLAS~\cite{ATLAS:2023bft}, and from approximately 10~$\MeVcc$ at 6500~$\MeVcc$ to 18~$\MeVcc$ at 7300~$\MeVcc$ for CMS~\cite{CMS:2023owd}. Consequently, the mass-resolution effect is negligible for LHCb but significant for both ATLAS and CMS. 
The CMS resolution~\cite{c_thesis_whw} is modeled by a double-Gaussian function, while the ATLAS resolution is derived from a second-order polynomial fit to the three available data points, as shown in Fig.~\ref{f_res}. 
Mass-resolution effects are incorporated by convolving a Gaussian function, $G(0;\sigma(m))$, with the signal component of the probability density function (PDF).

\subsection{Model I}
\begin{figure}[H]    
\centering    
\includegraphics[width=0.33\linewidth]{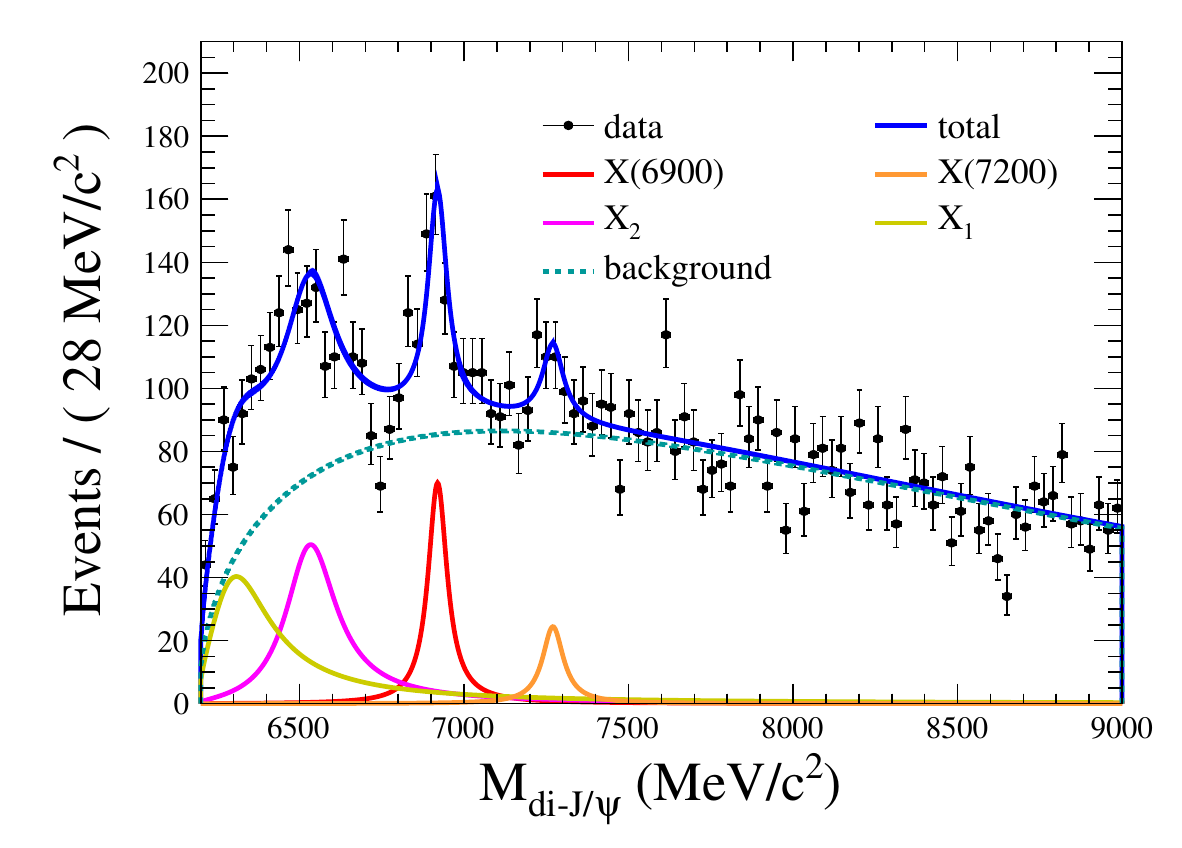}%
\includegraphics[width=0.33\linewidth]{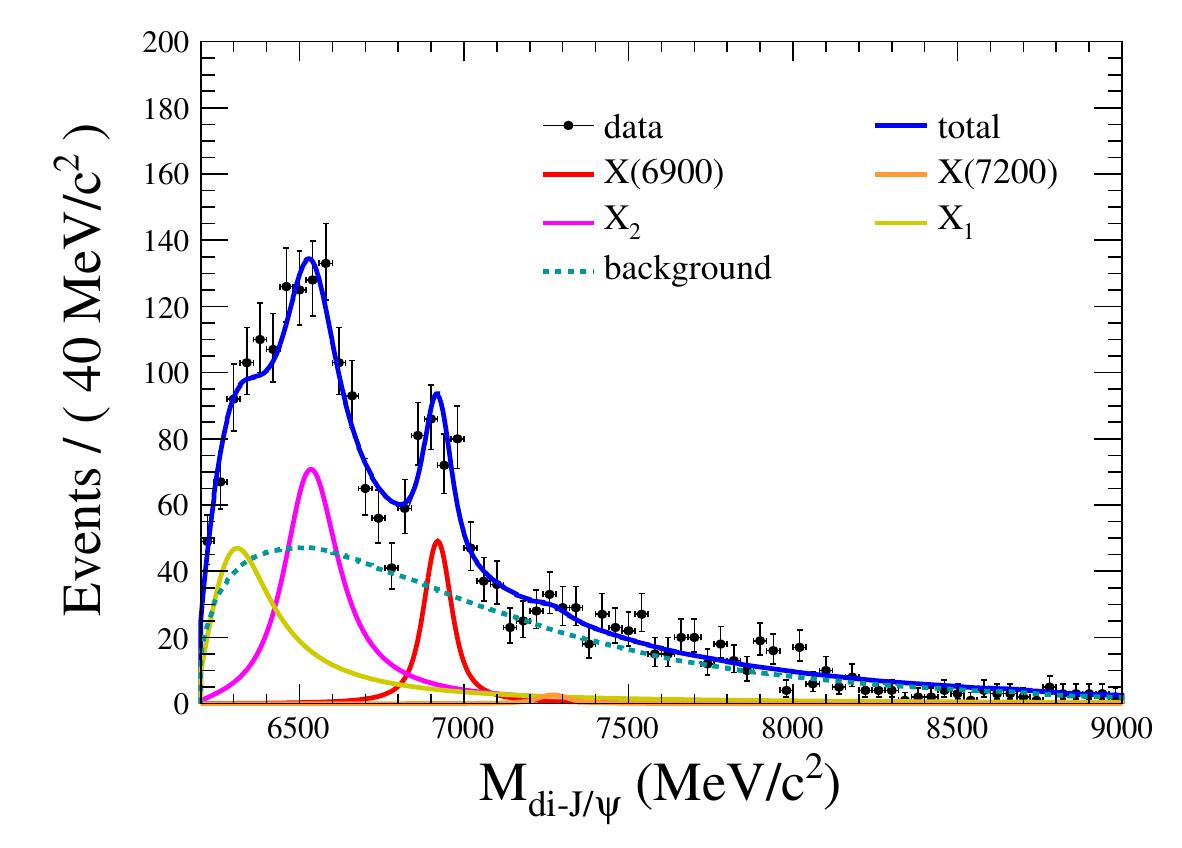}%
\includegraphics[width=0.33\linewidth]{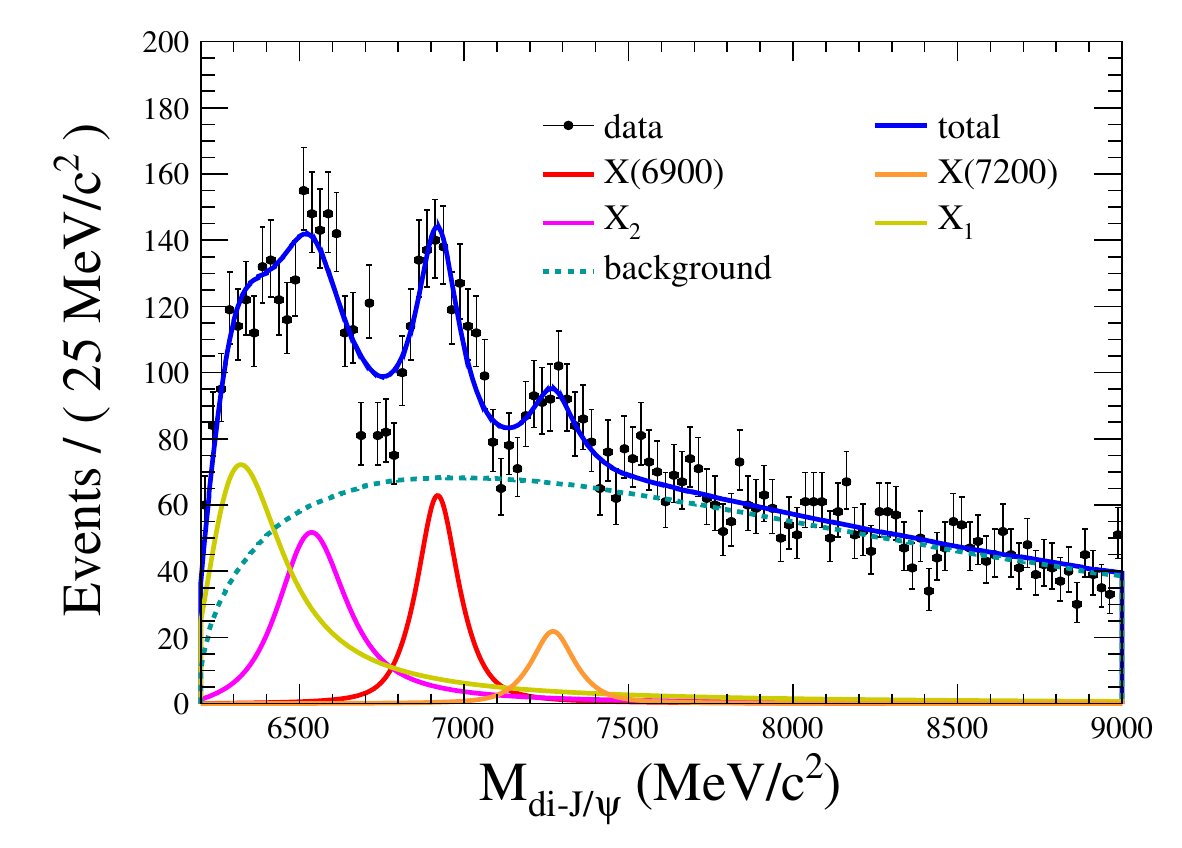}%

\caption{Fit results (Model~I) for the di-$\jpsi$ invariant-mass distributions from LHCb (left), ATLAS (middle), and CMS (right). The blue solid curves show the best fits; the individual contributions from the resonant components (solid) and the background (dotted) are also displayed.}    
\label{fig:result1}
\end{figure}

This model serves as the noninterfering baseline.
The total line shape is given by:
\begin{equation}
  f_{I}(m) = \left[\sum_{j} c_j \left| \text{BW}_j(m; M_j, \Gamma_j) \right|^2\right]\times \sqrt{1 - \frac{4M_{J/\psi}^2}{m^2}}\otimes G(0;\sigma(m))+B(m),
\end{equation}
where the index $j$ runs over $X_1$, $X_2$, $X(6900)$, and $X(7200)$; $c_j$ are free parameters that normalize each signal represented by a BW; and $\sqrt{1 - \frac{4M_{J/\psi}^2}{m^2}}$ is the phase-space factor. The background $B(m)$ accounts for contributions from SPS and DPS and is defined as $B(m) = \sqrt{1 - \frac{4M_{J/\psi}^2}{m^2}} \cdot f_{\rm SPS}(m) + f_{\rm DPS}(m)$. The fit result obtained under this hypothesis is shown in Fig.~\ref{fig:result1}. The extracted resonance parameters of the $X(6900)$ and $X(7200)$ are summarized in Table~\ref{t_LHCb_modelI}.

\subsection{Model II}
\begin{figure}[H]    
\centering    
\includegraphics[width=0.33\linewidth]{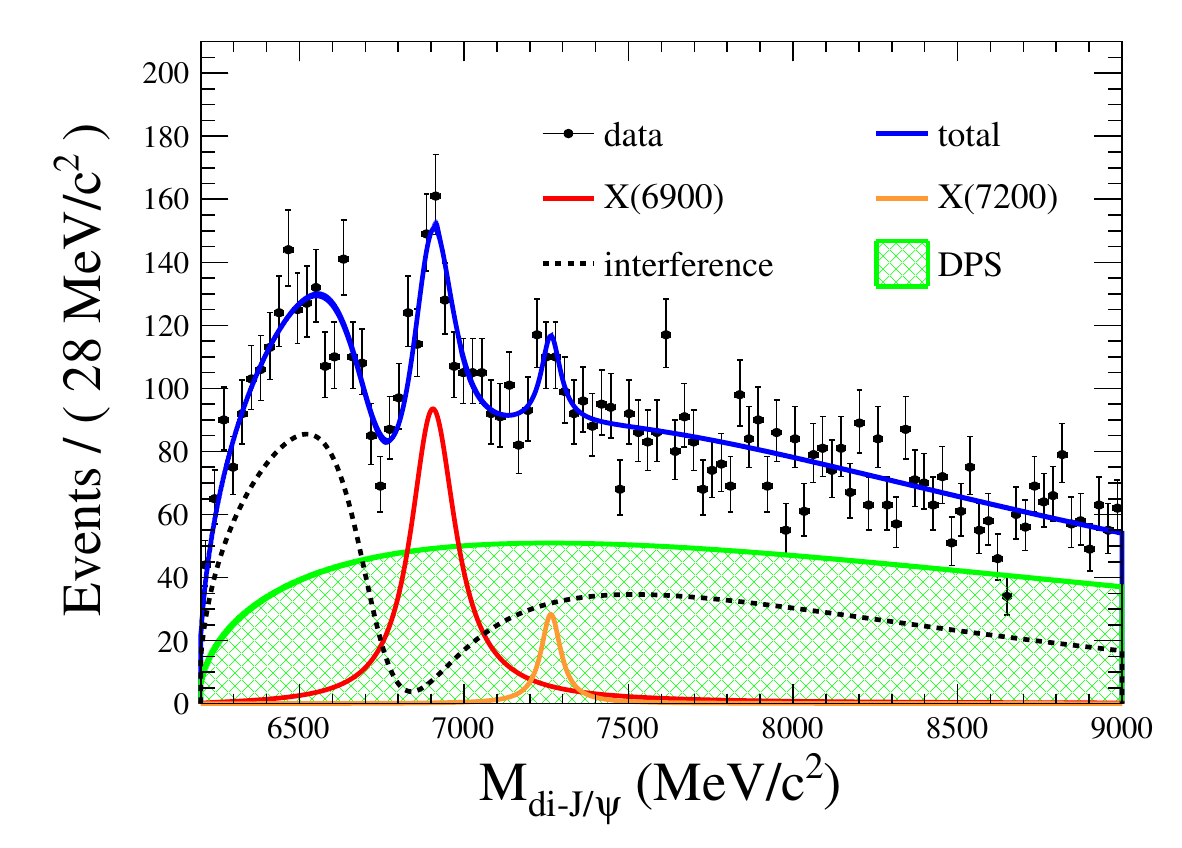}%
\includegraphics[width=0.33\linewidth]{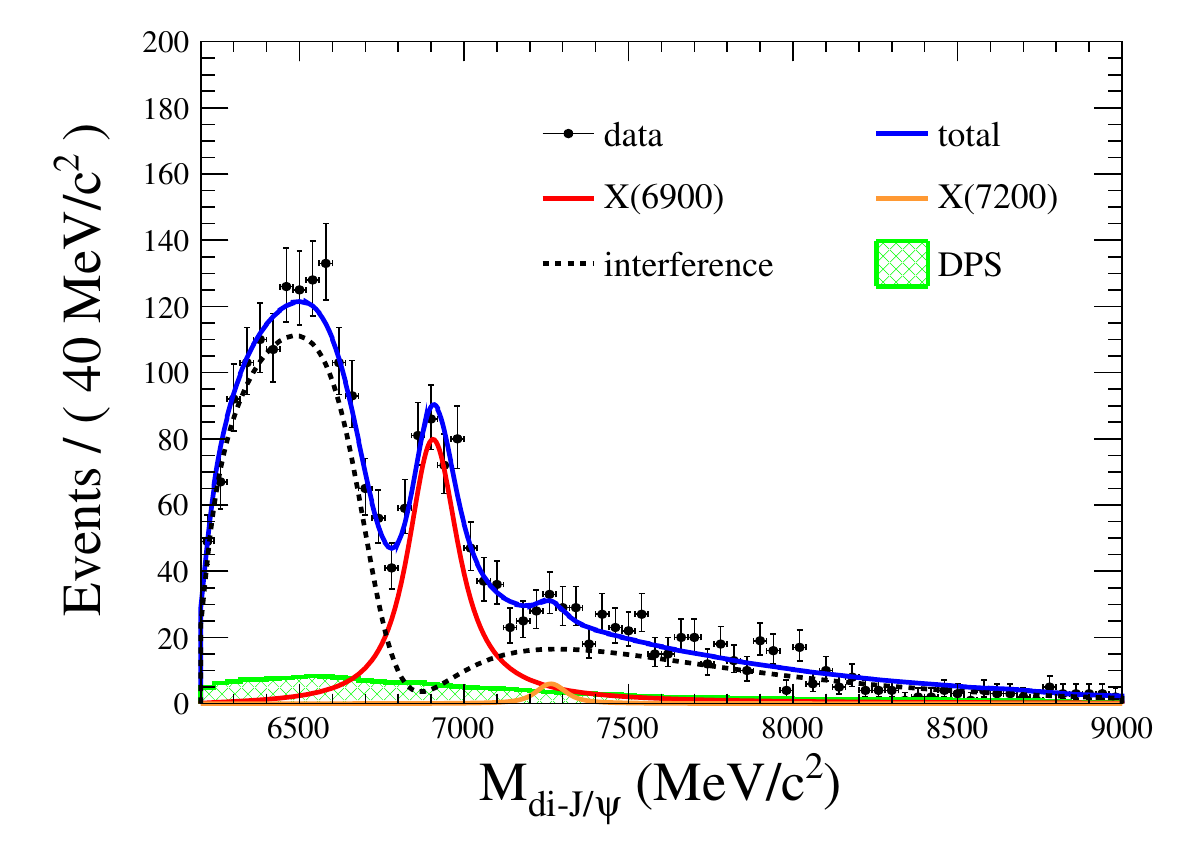}%
\includegraphics[width=0.33\linewidth]{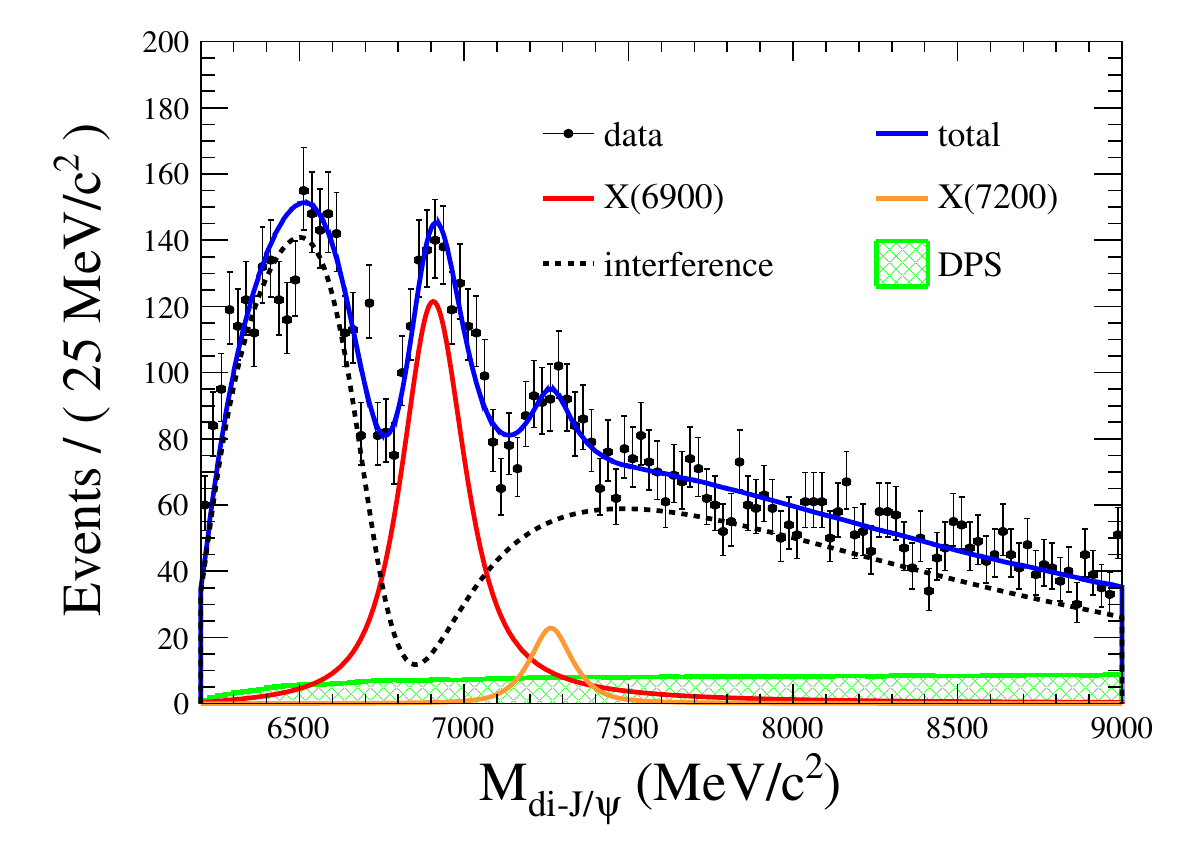}%
\caption{Fit results (Model~II) for the di-$\jpsi$ invariant-mass distributions measured by LHCb (left), ATLAS (middle), and CMS (right). The solid blue curves represent the best fit, and the individual contributions of the resonant components and the background are also shown.}    
\label{fig:result2}
\end{figure}

Following the LHCb Collaboration's treatment of interference, Model II couples the broad structure (described by an interfering BW) to the SPS background. By contrast, higher-mass states are modeled as isolated, noninterfering resonances:
\begin{equation}
\begin{aligned}
  f_{II}(m) &= \Bigg[ \left| \sqrt{f_{\rm SPS}(m)} + c\cdot e^{i\phi} \text{BW}(m;M,\Gamma) \right|^2 \\
&\quad + \left| \text{BW}_{X(6900)}(m) \right|^2 + \left| \text{BW}_{X(7200)}(m) \right|^2 \Bigg] \\
&\quad\times \sqrt{1 - \frac{4M_{J/\psi}^2}{m^2}}
 \otimes G(0;\sigma(m)) + f_{\rm DPS}(m)
\end{aligned}
\end{equation}
This configuration tests the hypothesis that the $X(6900)$ and $X(7200)$ are distinct, well-separated resonant states whose production phases are independent of the threshold enhancement. Figure~\ref{fig:result2} shows the fit results under this assumption, and Table~\ref{t_LHCb_modelI} summarizes the extracted resonance parameters for both structures.

\subsection{Model III}
\begin{figure}[H]    
\centering    
\includegraphics[width=0.33\linewidth]{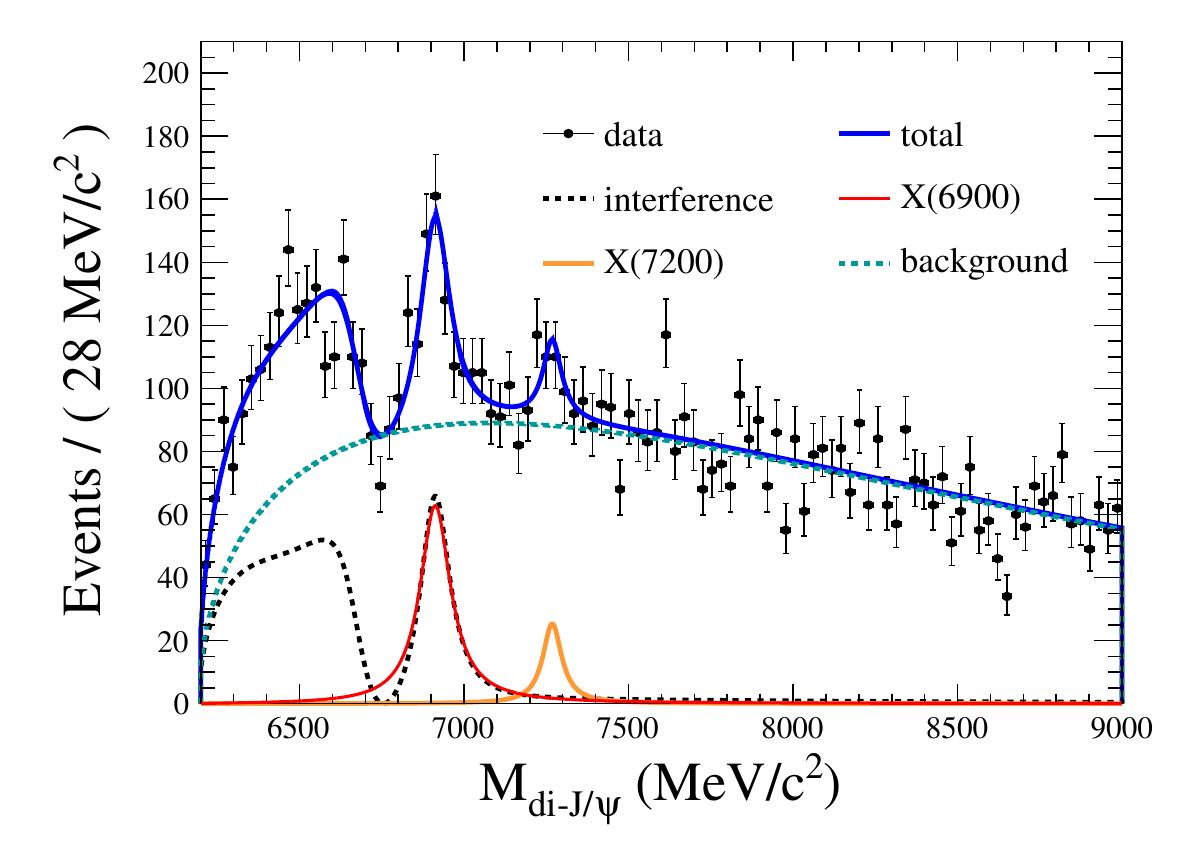}%
\includegraphics[width=0.33\linewidth]{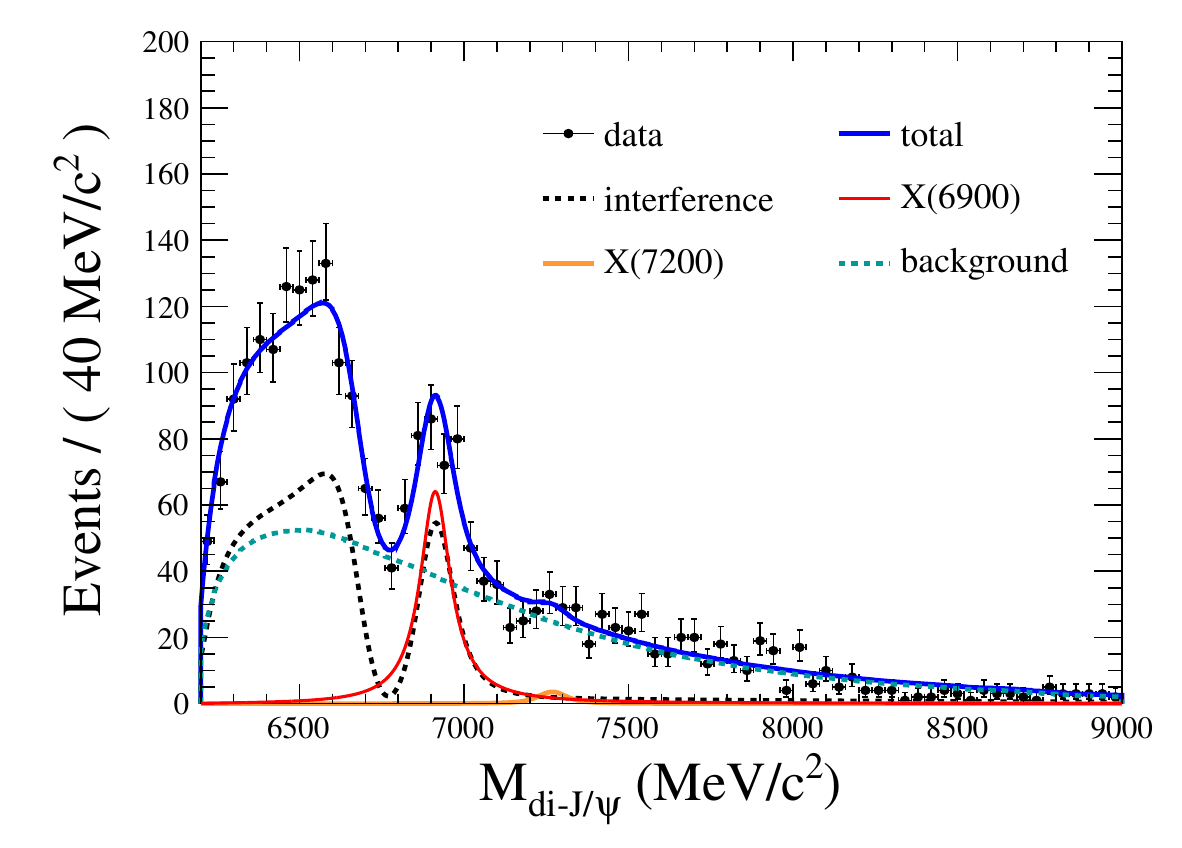}%
\includegraphics[width=0.33\linewidth]{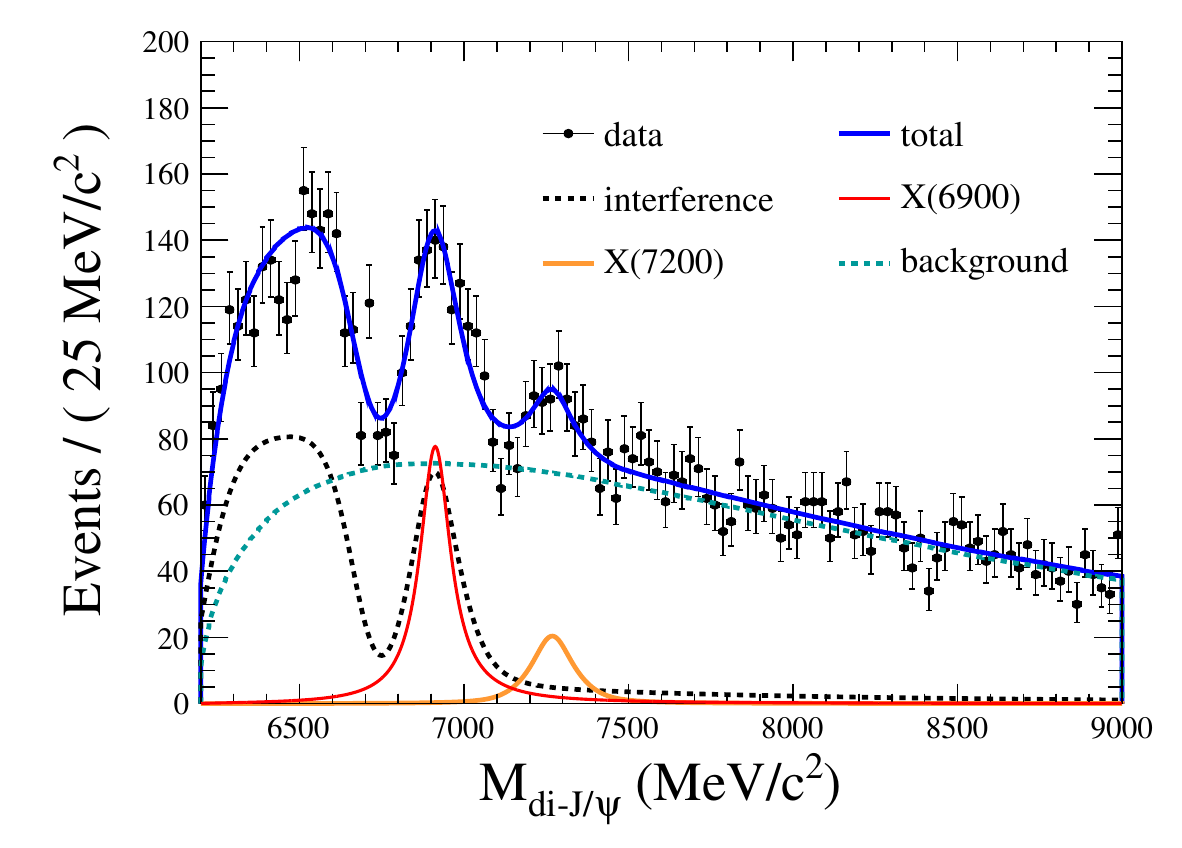}%
\caption{Fit results (Model~III) for the di-$\jpsi$ invariant-mass distributions from LHCb (left), CMS (middle), and ATLAS (right). The blue solid curves denote the best fit; individual contributions from the resonant components and the background are also shown.}    
\label{fig:result3}
\end{figure}

Model III investigates the scenario in which the three lower-mass structures ($X_1$, $X_2$, and $X(6900)$) share a common production mechanism (e.g., as $J^{PC} = 0^{++}$ states). In this interpretation, spectral dips arise from destructive interference among these coherently coupled resonances, whereas $X(7200)$ is treated as an incoherent component because its mass lies well above the threshold region, where nonresonant or multichannel effects are expected to dominate.
\begin{equation}
\begin{aligned}
  f_{III}(m) =&\left[\left| \sum_{j} c_j e^{i\phi_j} \text{BW}_j(m; M_j, \Gamma_j) \right|^2  + \left| \text{BW}_{X(7200)} \right|^2\right]\\
&\times \sqrt{1 - \frac{4M_{J/\psi}^2}{m^2}}\otimes G(0;\sigma(m)) + B(m),
\end{aligned}
\end{equation}
where the index $j$ runs over $X_1$, $X_2$, and $X(6900)$.
The fit result obtained under this hypothesis is shown in Fig.~\ref{fig:result3}. The extracted resonance parameters for $X(6900)$ and $X(7200)$ are summarized in Table~\ref{t_LHCb_modelI}.

\subsection{Model IV}
\begin{figure}[H]    
\centering    
\includegraphics[width=0.33\linewidth]{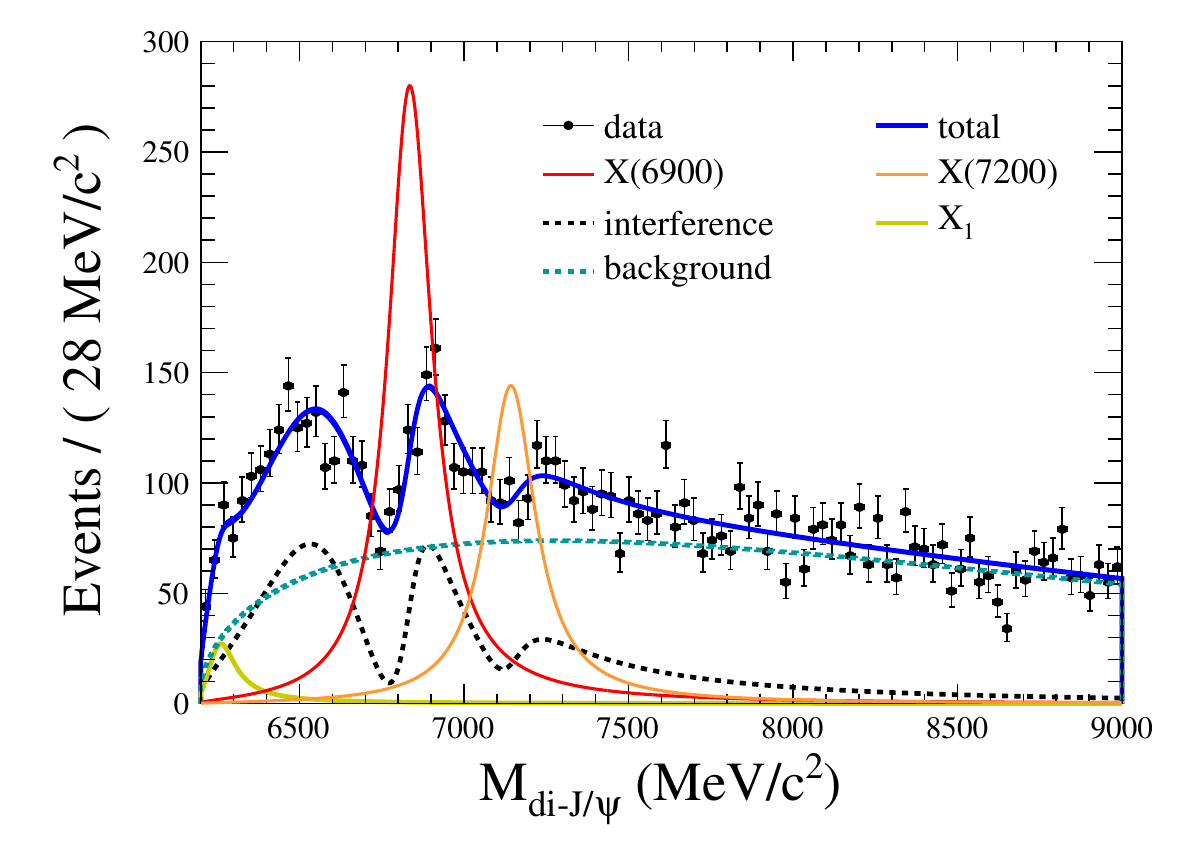}%
\includegraphics[width=0.33\linewidth]{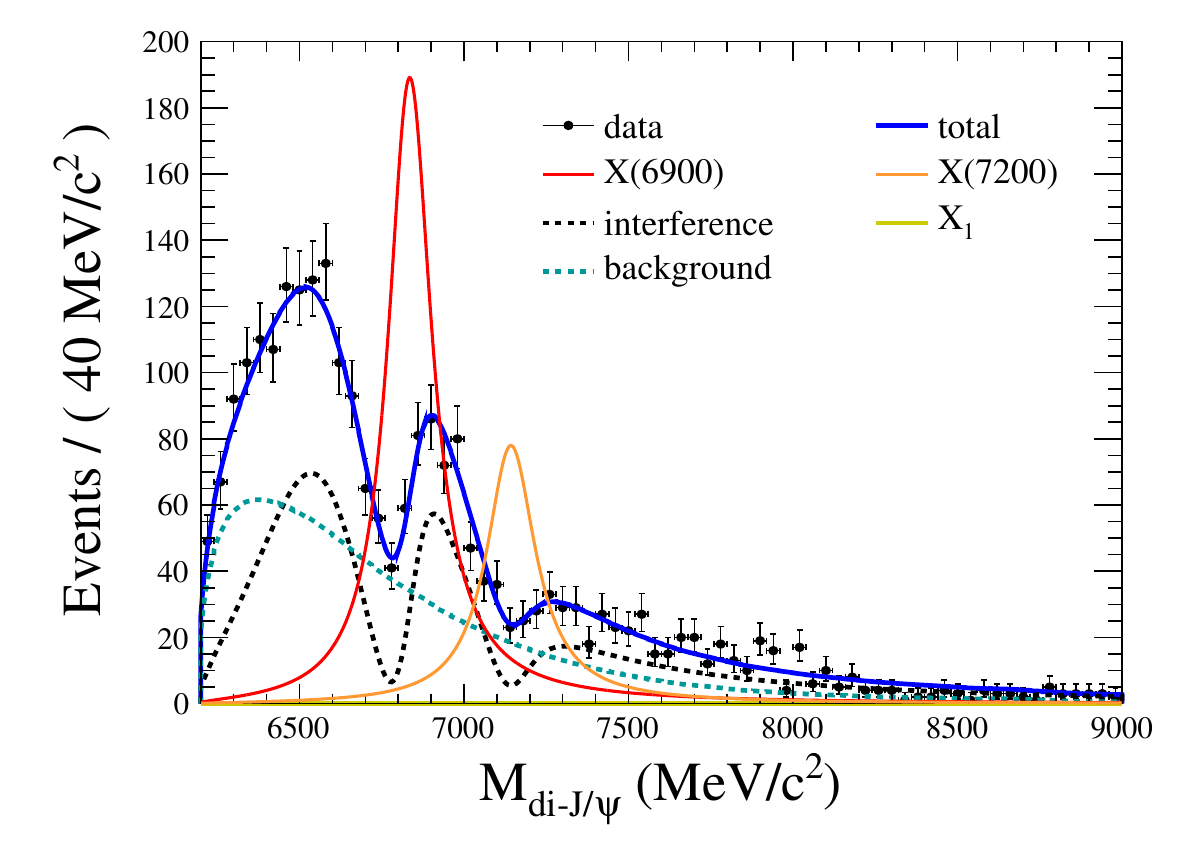}%
\includegraphics[width=0.33\linewidth]{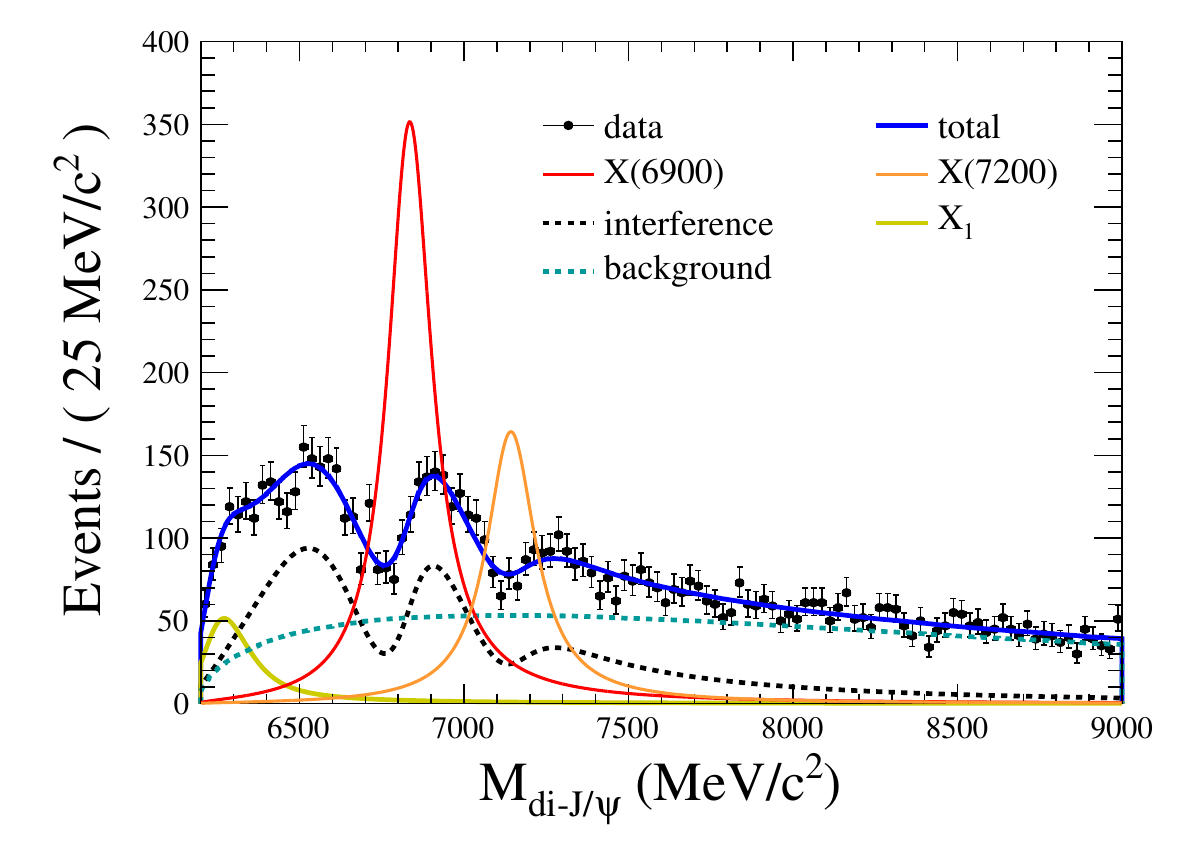}%
\caption{Fit results (Model~IV) for the di-$\jpsi$ invariant-mass distributions measured by LHCb (left), ATLAS (middle), and CMS (right). The solid blue curves represent the best fit, and the individual contributions of the resonant components and the background are also shown.}    
\label{fig:result4}
\end{figure}

Model~IV is motivated by the three-resonance interference model employed by CMS~\cite{CMS:2023owd}. In this scenario, $X_1$ is interpreted not as a genuine resonance, but as a threshold enhancement arising from nonresonant dynamics—such as coupled-channel effects, triangle singularities, or Pomeron exchange—which are difficult to model precisely because of large theoretical uncertainties in this energy region. Consequently, $X_1$ is treated as an incoherent background component, while $X_2$, $X(6900)$, and $X(7200)$ are allowed to interfere coherently as overlapping resonant states. This model tests whether $X(7200)$ is dynamically coupled to the $X(6900)$ structure.
\begin{equation}
\begin{aligned}
f_{IV}(m) = & \left[\left| \text{BW}_{X_{1}} \right|^2 + \left| \sum_{j} c_j\cdot e^{i\phi_j} \text{BW}_j(m; M_j, \Gamma_j )\right|^2 \right]\\
&\times \sqrt{1 - \frac{4M_{J/\psi}^2}{m^2}}\otimes  G(0;\sigma(m)) + B(m),
\end{aligned}
\end{equation}
where $j$ ranges over $X_2$, $X(6900)$, and $X(7200)$.
The fit result obtained under this hypothesis is shown in Fig.~\ref{fig:result4}. The extracted resonance parameters for the $X(6900)$ and $X(7200)$ structures are summarized in Table~\ref{t_LHCb_modelI}.

\begin{table}[H]    
\centering    
\caption{Fitted resonance parameters for the $X(6900)$ and $X(7200)$ structures are presented for four fit models. For each model, the mass ($M$), width ($\Gamma$), and statistical significance are reported. The $\chi^2/\mathrm{NDF}$ values quantify the goodness of fit for each model.}    
{\scriptsize
\begin{tabular}{l|c|c|c|c}    
\hline\hline    
Model & I & II & III & IV\\    
\hline         
$M_{X6900}$ ($\text{MeV}/c^2$)&$6919.27\pm2.90^{+3.5}_{-3.6}$&$6903.10\pm4.00^{+9.6}_{-8.9}$&$6911.61\pm7.60^{+7.3}_{-7.0}$&$6833.05\pm13.98^{+7.3}_{-7.0}$\\         
$\Gamma_{X6900}$ ($\text{MeV}$)& $70.25\pm8.48^{+19.8}_{-18.4}$&$165.22\pm14.00^{+41.4}_{-36.8}$&$112.74\pm14.36^{+28.6}_{-26.5}$&$161.31\pm19.43^{+28.6}_{-26.5}$\\

Significance &12.5$\sigma$ &15.6$\sigma$ &13.1$\sigma$ &14.1$\sigma$\\
\hline         
$M_{X7200}$ ($\text{MeV}/c^2$)&$7269.97\pm9.97\pm5.0$&$7263.69\pm8.59^{+41.8}_{-15.3}$&$7267.89\pm10.35^{+41.8}_{-15.3}$&  $7140.79\pm23.94^{+41.0}_{-15.0}$ \\         
$\Gamma_{X7200}$ ($\text{MeV}$)& $76.94\pm25.50\pm15.4$&$73.04\pm22.85^{+21.8}_{-19.6}$&$69.43\pm21.87^{+20.8}_{-18.6}$&$181.63\pm46.95^{+54.3}_{-48.7}$\\

Significance &4.1$\sigma$ & 6.5$\sigma$&3.7$\sigma$ &6.6$\sigma$\\
\hline
$\chi^2$/NDF& 321.00/259 = 1.24&299.13/261 = 1.15 &285.21/257 = 1.11 &268.87/257 = 1.05\\
\hline\hline    
\end{tabular}\label{t_LHCb_modelI}
}
\end{table}

Table~\ref{t_LHCb_modelI} summarizes the fitted resonance parameters of the $X(6900)$ and $X(7200)$ structures obtained under the four fit hypotheses. We observe clear variations in the extracted parameters across the different models, indicating a strong sensitivity of the fit results to the treatment of interference effects.  
Additionally, we account for the systematic uncertainties reported by all three collaborations~\cite{LHCb:2020bwg,ATLAS:2023bft,CMS:2023owd} using a weighted combination approach. We treat these uncertainties as mutually independent and combine them using inverse-variance weighting, with weights determined by the reported statistical uncertainties.

\begin{equation}
\sigma_{\rm comb} = \sqrt{\sum_i \left(w_i \sigma_{{\rm sys},i}\right)^2}, \quad w_i = \frac{1/\sigma_{{\rm stat},i}^2}{\sum_j 1/\sigma_{{\rm stat},j}^2},
\end{equation}
where $\sigma_{{\rm sys},i}$ and $\sigma_{{\rm stat},i}$ denote the systematic and statistical uncertainties for the $i^{\rm th}$ experiment, respectively.

\section{discussion and conclusion}
For the $X(6900)$ state, the extracted mass ranges from $6833$ to $6919~\MeVcc$ across the four models, with Model~IV yielding the lowest mass and a correspondingly larger width. This systematic shift indicates substantial interference effects in the $X(6900)$ line shape. The signal significance exceeds $12\sigma$ in all model configurations, demonstrating the structure's robustness against varying theoretical assumptions.

The $X(7200)$ structure displays greater model sensitivity, with both mass and width showing significant variations across different fit scenarios. Model~IV produces the most pronounced deviations---a substantial downward mass shift accompanied by enlarged uncertainties. The signal significance varies from $3.7\sigma$ to $6.6\sigma$, reflecting strong dependence on interference modeling. These observations necessitate a cautious interpretation of the physical origin of the $X(7200)$.

All four models provide an acceptable description of the data, with $\chi^{2}/\mathrm{NDF}$ values close to unity. Among them, Model~IV---which adopts the same three-resonance coherent interference scheme used by CMS~\cite{CMS:2023owd}---yields the best fit (with a significance of $6.6\sigma$) and is considered the most physically representative. Moreover, the fitted relative phase between $X(7200)$ and $X(6900)$ in Model~IV is $\phi = -0.79 \pm 0.24$~rad. This central value differs markedly from $0$ or $\pi$, which would be expected if one of the structures were purely nonresonant or if the contributions were incoherent. Instead, the nontrivial phase is consistent with coherent interference between two overlapping resonant amplitudes, providing further support for interpreting both $X(6900)$ and $X(7200)$ as genuine resonant states. Crucially, the background level in Model~IV is significantly reduced compared with Model~III---by approximately 14\% (LHCb), 19\% (ATLAS), and 21\% (CMS)---which directly enhances the signal-to-background ratio and thus the statistical significance.

The best combined fit (Model~IV) yields a mass of $6833 \pm 16\,\mathrm{MeV}/c^2$ for the $X(6900)$ and $7141 \pm 48\,\mathrm{MeV}/c^2$ for the $X(7200)$. These values are in good agreement with a range of theoretical approaches---including lattice QCD~\cite{Yang:2021hrb}, QCD sum rules~\cite{Chen:2016jxd}, and quark-model calculations~\cite{Liu:2021rtn,Tiwari:2021tmz,Giron:2020wpx}---which predict compact $c\bar{c}c\bar{c}$ states in the $6.8$--$7.3\,\mathrm{GeV}/c^2$ mass region. In particular, the mass of the $X(6900)$ is consistent with expectations for a $1P$-wave fully charmed tetraquark, while the $X(7200)$ may correspond to a radially excited ($2S$) state, as suggested in Ref.~\cite{Liu:2021rtn}. Furthermore, our results support the quark-delocalization color-screening mechanism proposed in Ref.~\cite{Jin:2020jfc}, which naturally accommodates multiple narrow tetraquark states in this energy region.

In summary, while the fully incoherent model serves as a phenomenological baseline for the di-$J/\psi$ spectrum, our analysis establishes that coherent interference effects are essential for a physically accurate description. These effects induce significant shifts in resonance parameters---particularly for the $X(6900)$---and enable robust characterization of the $X(7200)$. The results highlight the critical role of interference modeling in resolving overlapping all-charm resonances, offering new insights into tetraquark dynamics at LHC energies.

{\bf Acknowledgments}
This work was supported by the National Natural Science Foundation of China (NSFC) under Grant No.~12475077.

\bibliography{ref}

@article{Gell-Mann:1964ewy,
    author = "Gell-Mann, Murray",
    title = "{A Schematic Model of Baryons and Mesons}",
    doi = "10.1016/S0031-9163(64)92001-3",
    journal = "Phys. Lett.",
    volume = "8",
    pages = "214--215",
    year = "1964"
}

@inbook{Zweig:1964jf,
    author = "Zweig, G.",
    editor = "Lichtenberg, D. B. and Rosen, Simon Peter",
    title = "{An SU(3) model for strong interaction symmetry and its breaking. Version 2}",
    booktitle = "{DEVELOPMENTS IN THE QUARK THEORY OF HADRONS. VOL. 1. 1964 - 1978}",
    reportNumber = "CERN-TH-412, NP-14146, PRINT-64-170",
    doi = "10.17181/CERN-TH-412",
    pages = "22--101",
    month = "2",
    year = "1964"
}

@article{Belle:2003nnu,
    author = "Choi, S. K. and others",
    collaboration = "Belle",
    title = "{Observation of a narrow charmonium-like state in exclusive $B^\pm \to K^\pm \pi^+ \pi^- J/\psi$ decays}",
    eprint = "hep-ex/0309032",
    archivePrefix = "arXiv",
    doi = "10.1103/PhysRevLett.91.262001",
    journal = "Phys. Rev. Lett.",
    volume = "91",
    pages = "262001",
    year = "2003"
}

@article{LHCb:2015yax,
    author = "Aaij, Roel and others",
    collaboration = "LHCb",
    title = "{Observation of $J/\psi p$ Resonances Consistent with Pentaquark States in $\Lambda_b^0 \to J/\psi K^- p$ Decays}",
    eprint = "1507.03414",
    archivePrefix = "arXiv",
    primaryClass = "hep-ex",
    reportNumber = "CERN-PH-EP-2015-153, LHCB-PAPER-2015-029",
    doi = "10.1103/PhysRevLett.115.072001",
    journal = "Phys. Rev. Lett.",
    volume = "115",
    pages = "072001",
    year = "2015"
}

@article{LHCb:2019kea,
    author = "Aaij, Roel and others",
    collaboration = "LHCb",
    title = "{Observation of a narrow pentaquark state, $P_c(4312)^+$, and of two-peak structure of the $P_c(4450)^+$}",
    eprint = "1904.03947",
    archivePrefix = "arXiv",
    primaryClass = "hep-ex",
    reportNumber = "LHCb-PAPER-2019-014 CERN-EP-2019-058",
    doi = "10.1103/PhysRevLett.122.222001",
    journal = "Phys. Rev. Lett.",
    volume = "122",
    number = "22",
    pages = "222001",
    year = "2019"
}

@article{LHCb:2020bwg,
    author = "Aaij, Roel and others",
    collaboration = "LHCb",
    title = "{Observation of structure in the $J /\psi$ -pair mass spectrum}",
    eprint = "2006.16957",
    archivePrefix = "arXiv",
    primaryClass = "hep-ex",
    reportNumber = "CERN-EP-2020-115, LHCb-PAPER-2020-011",
    doi = "10.1016/j.scib.2020.08.032",
    journal = "Sci. Bull.",
    volume = "65",
    number = "23",
    pages = "1983--1993",
    year = "2020"
}

@article{Dong:2020nwy,
    author = "Dong, Xiang-Kun and Baru, Vadim and Guo, Feng-Kun and Hanhart, Christoph and Nefediev, Alexey",
    title = "{Coupled-Channel Interpretation of the LHCb Double-~$J/\psi$~Spectrum and Hints of a New State Near the~ $J/\psi J/\psi$~~Threshold}",
    eprint = "2009.07795",
    archivePrefix = "arXiv",
    primaryClass = "hep-ph",
    doi = "10.1103/PhysRevLett.127.119901",
    journal = "Phys. Rev. Lett.",
    volume = "126",
    number = "13",
    pages = "132001",
    year = "2021",
    note = "[Erratum: Phys.Rev.Lett. 127, 119901 (2021)]"
}

@article{CMS:2023owd,
    author = "Hayrapetyan, Aram and others",
    collaboration = "CMS",
    title = "{New Structures in the J/{\ensuremath{\psi}}J/{\ensuremath{\psi}} Mass Spectrum in Proton-Proton Collisions at s=13{\,}{\,}TeV}",
    eprint = "2306.07164",
    archivePrefix = "arXiv",
    primaryClass = "hep-ex",
    reportNumber = "CMS-BPH-21-003, CERN-EP-2023-109",
    doi = "10.1103/PhysRevLett.132.111901",
    journal = "Phys. Rev. Lett.",
    volume = "132",
    number = "11",
    pages = "111901",
    year = "2024"
}

@article{ATLAS:2023bft,
    author = "Aad, Georges and others",
    collaboration = "ATLAS",
    title = "{Observation of an Excess of Dicharmonium Events in the Four-Muon Final State with the ATLAS Detector}",
    eprint = "2304.08962",
    archivePrefix = "arXiv",
    primaryClass = "hep-ex",
    reportNumber = "CERN-EP-2023-035",
    doi = "10.1103/PhysRevLett.131.151902",
    journal = "Phys. Rev. Lett.",
    volume = "131",
    number = "15",
    pages = "151902",
    year = "2023"
}

@article{Wang:2020wrp,
    author = "Wang, Jun-Zhang and Chen, Dian-Yong and Liu, Xiang and Matsuki, Takayuki",
    title = "{Producing fully charm structures in the $J/\psi$ -pair invariant mass spectrum}",
    eprint = "2008.07430",
    archivePrefix = "arXiv",
    primaryClass = "hep-ph",
    doi = "10.1103/PhysRevD.103.L071503",
    journal = "Phys. Rev. D",
    volume = "103",
    number = "7",
    pages = "071503",
    year = "2021"
}

@article{Chen:2020xwe,
    author = "Chen, Hua-Xing and Chen, Wei and Liu, Xiang and Zhu, Shi-Lin",
    title = "{Strong decays of fully-charm tetraquarks into di-charmonia}",
    eprint = "2006.16027",
    archivePrefix = "arXiv",
    primaryClass = "hep-ph",
    doi = "10.1016/j.scib.2020.08.038",
    journal = "Sci. Bull.",
    volume = "65",
    pages = "1994--2000",
    year = "2020"
}

@article{Bedolla:2019zwg,
    author = "Bedolla, M. A. and Ferretti, J. and Roberts, C. D. and Santopinto, E.",
    title = "{Spectrum of fully-heavy tetraquarks from a diquark+antidiquark perspective}",
    eprint = "1911.00960",
    archivePrefix = "arXiv",
    primaryClass = "hep-ph",
    reportNumber = "NJU-INP 009/19",
    doi = "10.1140/epjc/s10052-020-08579-3",
    journal = "Eur. Phys. J. C",
    volume = "80",
    number = "11",
    pages = "1004",
    year = "2020"
}

@article{Jin:2020jfc,
    author = "Jin, Xin and Xue, Yaoyao and Huang, Hongxia and Ping, Jialun",
    title = "{Full-heavy tetraquarks in constituent quark models}",
    eprint = "2006.13745",
    archivePrefix = "arXiv",
    primaryClass = "hep-ph",
    doi = "10.1140/epjc/s10052-020-08650-z",
    journal = "Eur. Phys. J. C",
    volume = "80",
    number = "11",
    pages = "1083",
    year = "2020"
}

@article{liu:2020eha,
    author = "liu, Ming-Sheng and Liu, Feng-Xiao and Zhong, Xian-Hui and Zhao, Qiang",
    title = "{Fully heavy tetraquark states and their evidences in LHC observations}",
    eprint = "2006.11952",
    archivePrefix = "arXiv",
    primaryClass = "hep-ph",
    doi = "10.1103/PhysRevD.109.076017",
    journal = "Phys. Rev. D",
    volume = "109",
    number = "7",
    pages = "076017",
    year = "2024"
}

@article{Liu:2021rtn,
    author = "Liu, Feng-Xiao and Liu, Ming-Sheng and Zhong, Xian-Hui and Zhao, Qiang",
    title = "{Higher mass spectra of the fully-charmed and fully-bottom tetraquarks}",
    eprint = "2110.09052",
    archivePrefix = "arXiv",
    primaryClass = "hep-ph",
    doi = "10.1103/PhysRevD.104.116029",
    journal = "Phys. Rev. D",
    volume = "104",
    number = "11",
    pages = "116029",
    year = "2021"
}

@article{Zhu:2020xni,
    author = "Zhu, Ruilin",
    title = "{Fully-heavy tetraquark spectra and production at hadron colliders}",
    eprint = "2010.09082",
    archivePrefix = "arXiv",
    primaryClass = "hep-ph",
    doi = "10.1016/j.nuclphysb.2021.115393",
    journal = "Nucl. Phys. B",
    volume = "966",
    pages = "115393",
    year = "2021"
}

@article{Giron:2020wpx,
    author = "Giron, Jesse F. and Lebed, Richard F.",
    title = "{Simple spectrum of $c\bar c c\bar c$ states in the dynamical diquark model}",
    eprint = "2008.01631",
    archivePrefix = "arXiv",
    primaryClass = "hep-ph",
    doi = "10.1103/PhysRevD.102.074003",
    journal = "Phys. Rev. D",
    volume = "102",
    number = "7",
    pages = "074003",
    year = "2020"
}

@article{Iwasaki:1976cn,
    author = "Iwasaki, Y.",
    title = "{Is a State c anti-c c anti-c Found at 6.0-GeV?}",
    reportNumber = "UTHEP-7",
    doi = "10.1103/PhysRevLett.36.1266",
    journal = "Phys. Rev. Lett.",
    volume = "36",
    pages = "1266",
    year = "1976"
}

@article{Chao:1980dv,
    author = "Chao, Kuang-Ta",
    title = "{The (cc) - ($\bar{cc}$) (Diquark - Anti-Diquark) States in $e^+ e^-$ Annihilation}",
    reportNumber = "FERMILAB-PUB-80-070-THY, FERMILAB-PUB-80-070-T",
    doi = "10.1007/BF01431564",
    journal = "Z. Phys. C",
    volume = "7",
    pages = "317",
    year = "1981"
}

@article{Berezhnoy:2011xn,
    author = "Berezhnoy, A. V. and Luchinsky, A. V. and Novoselov, A. A.",
    title = "{Tetraquarks Composed of 4 Heavy Quarks}",
    eprint = "1111.1867",
    archivePrefix = "arXiv",
    primaryClass = "hep-ph",
    doi = "10.1103/PhysRevD.86.034004",
    journal = "Phys. Rev. D",
    volume = "86",
    pages = "034004",
    year = "2012"
}

@article{Wu:2016vtq,
    author = "Wu, Jing and Liu, Yan-Rui and Chen, Kan and Liu, Xiang and Zhu, Shi-Lin",
    title = "{Heavy-flavored tetraquark states with the $QQ\bar{Q}\bar{Q}$ configuration}",
    eprint = "1605.01134",
    archivePrefix = "arXiv",
    primaryClass = "hep-ph",
    doi = "10.1103/PhysRevD.97.094015",
    journal = "Phys. Rev. D",
    volume = "97",
    number = "9",
    pages = "094015",
    year = "2018"
}

@article{Bai:2016int,
    author = "Bai, Yang and Lu, Sida and Osborne, James",
    title = "{Beauty-full Tetraquarks}",
    eprint = "1612.00012",
    archivePrefix = "arXiv",
    primaryClass = "hep-ph",
    doi = "10.1016/j.physletb.2019.134930",
    journal = "Phys. Lett. B",
    volume = "798",
    pages = "134930",
    year = "2019"
}

@article{Wang:2017jtz,
    author = "Wang, Zhi-Gang",
    title = "{Analysis of the $QQ\bar{Q}\bar{Q}$ tetraquark states with QCD sum rules}",
    eprint = "1701.04285",
    archivePrefix = "arXiv",
    primaryClass = "hep-ph",
    doi = "10.1140/epjc/s10052-017-4997-0",
    journal = "Eur. Phys. J. C",
    volume = "77",
    number = "7",
    pages = "432",
    year = "2017"
}

@article{Anwar:2017toa,
    author = "Anwar, Muhammad Naeem and Ferretti, Jacopo and Guo, Feng-Kun and Santopinto, Elena and Zou, Bing-Song",
    title = "{Spectroscopy and decays of the fully-heavy tetraquarks}",
    eprint = "1710.02540",
    archivePrefix = "arXiv",
    primaryClass = "hep-ph",
    doi = "10.1140/epjc/s10052-018-6073-9",
    journal = "Eur. Phys. J. C",
    volume = "78",
    number = "8",
    pages = "647",
    year = "2018"
}

@article{Richard:2017vry,
    author = "Richard, Jean-Marc and Valcarce, Alfredo and Vijande, Javier",
    title = "{String dynamics and metastability of all-heavy tetraquarks}",
    eprint = "1703.00783",
    archivePrefix = "arXiv",
    primaryClass = "hep-ph",
    doi = "10.1103/PhysRevD.95.054019",
    journal = "Phys. Rev. D",
    volume = "95",
    number = "5",
    pages = "054019",
    year = "2017"
}

@article{Esposito:2018cwh,
    author = "Esposito, Angelo and Polosa, Antonio D.",
    title = "{A $bb\bar b\bar b$ di-bottomonium at the LHC?}",
    eprint = "1807.06040",
    archivePrefix = "arXiv",
    primaryClass = "hep-ph",
    doi = "10.1140/epjc/s10052-018-6269-z",
    journal = "Eur. Phys. J. C",
    volume = "78",
    number = "9",
    pages = "782",
    year = "2018"
}

@article{Karliner:2016zzc,
    author = "Karliner, Marek and Nussinov, Shmuel and Rosner, Jonathan L.",
    title = "{$Q Q \bar Q \bar Q$ states: masses, production, and decays}",
    eprint = "1611.00348",
    archivePrefix = "arXiv",
    primaryClass = "hep-ph",
    reportNumber = "EFI-16-22, TAUP-3012-16",
    doi = "10.1103/PhysRevD.95.034011",
    journal = "Phys. Rev. D",
    volume = "95",
    number = "3",
    pages = "034011",
    year = "2017"
}

@article{Liu:2019zuc,
    author = {Liu, Ming-Sheng and L{\"u}, Qi-Fang and Zhong, Xian-Hui and Zhao, Qiang},
    title = "{All-heavy tetraquarks}",
    eprint = "1901.02564",
    archivePrefix = "arXiv",
    primaryClass = "hep-ph",
    doi = "10.1103/PhysRevD.100.016006",
    journal = "Phys. Rev. D",
    volume = "100",
    number = "1",
    pages = "016006",
    year = "2019"
}

@article{Chen:2016jxd,
    author = "Chen, Wei and Chen, Hua-Xing and Liu, Xiang and Steele, T. G. and Zhu, Shi-Lin",
    title = "{Hunting for exotic doubly hidden-charm/bottom tetraquark states}",
    eprint = "1605.01647",
    archivePrefix = "arXiv",
    primaryClass = "hep-ph",
    doi = "10.1016/j.physletb.2017.08.034",
    journal = "Phys. Lett. B",
    volume = "773",
    pages = "247--251",
    year = "2017"
}

@article{Wang:2019rdo,
    author = "Wang, Guang-Juan and Meng, Lu and Zhu, Shi-Lin",
    title = "{Spectrum of the fully-heavy tetraquark state $QQ\bar Q' \bar Q'$}",
    eprint = "1907.05177",
    archivePrefix = "arXiv",
    primaryClass = "hep-ph",
    doi = "10.1103/PhysRevD.100.096013",
    journal = "Phys. Rev. D",
    volume = "100",
    number = "9",
    pages = "096013",
    year = "2019"
}

@article{Gong:2020bmg,
    author = "Gong, Chang and Du, Meng-Chuan and Zhao, Qiang and Zhong, Xian-Hui and Zhou, Bin",
    title = "{Nature of X(6900) and its production mechanism at LHCb}",
    eprint = "2011.11374",
    archivePrefix = "arXiv",
    primaryClass = "hep-ph",
    doi = "10.1016/j.physletb.2021.136794",
    journal = "Phys. Lett. B",
    volume = "824",
    pages = "136794",
    year = "2022"
}

@article{Wan:2020fsk,
    author = "Wan, Bing-Dong and Qiao, Cong-Feng",
    title = "{Gluonic tetracharm configuration of $X (6900)$}",
    eprint = "2012.00454",
    archivePrefix = "arXiv",
    primaryClass = "hep-ph",
    doi = "10.1016/j.physletb.2021.136339",
    journal = "Phys. Lett. B",
    volume = "817",
    pages = "136339",
    year = "2021"
}

@article{Wang:2020tpt,
    author = "Wang, Jun-Zhang and Liu, Xiang and Matsuki, Takayuki",
    title = "{Fully-heavy structures in the invariant mass spectrum of $J/\psi \psi(3686)$, $J/\psi \psi(3770)$, $\psi(3686) \psi(3686)$, and $J/\psi \Upsilon(1S)$ at hadron colliders}",
    eprint = "2012.03281",
    archivePrefix = "arXiv",
    primaryClass = "hep-ph",
    doi = "10.1016/j.physletb.2021.136209",
    journal = "Phys. Lett. B",
    volume = "816",
    pages = "136209",
    year = "2021"
}

@phdthesis{c_thesis_whw,
  author  = "Wen, Hongwei",
  title   = "{Study of Near-Threshold Structures in the $\jpsi\jpsi$ Mass Spectrum at CMS}",
  school  = "Nanjing Normal University",
  year    = "2022",
  address = "Nanjing, China",
  month   = "August",
  note    = "",
  url     = ""
}

@article{CMS:2025fpt,
    author = "Hayrapetyan, Aram and others",
    collaboration = "CMS",
    title = "{Determination of the spin and parity of all-charm tetraquarks}",
    eprint = "2506.07944",
    archivePrefix = "arXiv",
    primaryClass = "hep-ex",
    reportNumber = "CMS-BPH-24-002, CERN-EP-2025-118",
    doi = "10.1038/s41586-025-09711-7",
    journal = "Nature",
    volume = "648",
    number = "8092",
    pages = "58--63",
    year = "2025"
}

@article{Yang:2021hrb,
    author = "Yang, Gang and Ping, Jialun and Segovia, Jorge",
    title = "{Exotic resonances of fully-heavy tetraquarks in a lattice-QCD insipired quark model}",
    eprint = "2104.08814",
    archivePrefix = "arXiv",
    primaryClass = "hep-ph",
    doi = "10.1103/PhysRevD.104.014006",
    journal = "Phys. Rev. D",
    volume = "104",
    number = "1",
    pages = "014006",
    year = "2021"
}

@article{Tiwari:2021tmz,
    author = "Tiwari, Rohit and Rathaud, D. P. and Rai, A. K.",
    title = "{Spectroscopy of all charm tetraquark states}",
    eprint = "2108.04017",
    archivePrefix = "arXiv",
    primaryClass = "hep-ph",
    doi = "10.1007/s12648-022-02427-8",
    journal = "Indian J. Phys.",
    volume = "97",
    number = "3",
    pages = "943--954",
    year = "2023"
}

@article{Guo:2019twa,
    author = "Guo, Feng-Kun and Liu, Xiao-Hai and Sakai, Shuntaro",
    title = "{Threshold cusps and triangle singularities in hadronic reactions}",
    eprint = "1912.07030",
    archivePrefix = "arXiv",
    primaryClass = "hep-ph",
    doi = "10.1016/j.ppnp.2020.103757",
    journal = "Prog. Part. Nucl. Phys.",
    volume = "112",
    pages = "103757",
    year = "2020"
}

@article{Wang:2025hex,
    author = "Wang, Yefan and Zhu, Ruilin",
    title = "{Fully charm tetraquark production at hadronic collisions with gluon radiation effects}",
    eprint = "2510.02085",
    archivePrefix = "arXiv",
    primaryClass = "hep-ph",
    month = "10",
    journal="",
    year = "2025"
}
\end{document}